\documentclass[aps,prl,twocolumn,showpacs]{revtex4-1}
\usepackage{epsfig}
\usepackage{graphicx}
\usepackage{bm}
\usepackage{amssymb}
\usepackage{amsmath}
\usepackage{amsthm}
\begin{document}
\title{Dynamic hysteresis in cyclic deformation of
crystalline solids}
\author{Lasse Laurson and Mikko J. Alava}
\affiliation{COMP Centre of Excellence, Department of Applied
Physics, Aalto University, PO Box 14100, 00076 Aalto, Espoo, Finland.}
\begin{abstract}
The hysteresis or internal friction in the deformation
of crystalline solids stressed cyclically is studied
from the viewpoint of collective dislocation dynamics.
Stress-controlled simulations of a dislocation
dynamics model at various loading frequencies and
amplitudes are performed to study the stress - {\it strain
rate} hysteresis. The hysteresis loop areas exhibit a maximum
at a characteristic frequency and a power law
frequency dependence in the low frequency limit, with
the power law exponent exhibiting two regimes, corresponding
to the jammed and the yielding/moving phases of the system, respectively.
The first of these phases exhibits non-trivial
critical-like viscoelastic dynamics, crossing over to
intermittent viscoplastic deformation for higher stress
amplitudes.
\end{abstract}
\maketitle

The response of interacting many-body systems to
oscillating external fields is an old problem in
physics, with many applications in materials science
and engineering. In general, due to the competing time
scales of the internal relaxation and the external
perturbation the response will generally be out of phase
with respect to the external field \cite{CHA-99}.
This gives rise to a dynamic hysteresis loop with an
area depending on the driving frequency and amplitude.
In a magnet driven by an oscillating
magnetic field $h(t)$, the magnetization $m(t)$
lagging behind the field leads to a non-vanishing
hysteresis loop area $A=\oint m dh$ \cite{MAGNETIC}.
Hysteresis as such is a very general phenomenon, and has been
studied in many contexts as from the mechanical
response of materials \cite{MECH}, to electronics \cite{EL},
cell biology \cite{CELL}, neurobiology \cite{NEURO} and
quantum systems \cite{QM}.

Mechanical dissipation or internal friction is one
manifestation of the dynamics of dislocations in crystalline
solids. Stress-strain hysteresis, in stress or strain
controlled experiments \cite{MECH}, is related via the
hysteresis loop area to the energy dissipated per cycle.
Since dislocations are line-like objects, internal friction
has also been described microscopically by the back-and-forth
dissipative motion of individual dislocation segments
\cite{DAN-97}. However, plastic, irreversible deformation
has been shown over the last decade to be a highly
co-operative process with avalanche dynamics and long-range
spatio-temporal correlations \cite{DISLO}.
Even the simplest dislocation dynamics models - which nevertheless
describe to a large degree some real materials - demonstrate
phenomena like a yielding/jamming transition at an
applied stress $\sigma = \sigma_c$ separating a phase
with frozen dislocations from a moving phase with a
stress-dependent average strain rate - the {\it order parameter} of
the transition \cite{MIG-02,MIG-08,LAU-10}. This is in analogy
to systems exhibiting criticality due to a depinning transition -
separating in the adiabatic, thermodynamic limit frozen and active
states, with the order parameter given by the average velocity -
such as interfaces in random media \cite{GLA-03} and vortices in
type-II superconductors \cite{MET-98}.

In this Letter we consider the dynamic {\it strain rate}
hysteresis of dislocation assemblies from the viewpoint of
collective dislocation dynamics. The important aspects are
i) the various behaviors in the phase diagram, characterized
by the amplitude and the frequency of the external
driving, ii) the collective phenomena that underlie the
observations from the simulations, and iii) the theoretical
and experimental implications of our results. We discuss the
scaling of the hysteresis, and link it to a picture related
to depinning transitions.
Recent theoretical ideas suggest that due to
the long-range dislocation stress fields this transition
should be described by the mean field depinning transition
\cite{ZAI-05,DAH-09}. However, our results point out that
this simple picture is incomplete, calling for novel theoretical
ideas to properly describe the glassy, critical-like dynamics
observed in the jammed phase of the system.

Dislocation physics has been recently studied with many
simplified models from discrete dislocation dynamics
\cite{DDD2D1,DDD2D2,DDD3D} to phase field \cite{PF} and automaton
models \cite{SAL-11}. Here,
we consider the stress-controlled hysteretic dynamics within a
two-dimensional discrete dislocations dynamics model \cite{DDD2D1}.
Such a model captures many of the interesting aspects of real
crystal plasticity, including the scale free distribution of avalanches
of plastic deformation \cite{MIG-01}, as well as an Andrade primary
creep law \cite{MIG-02,MIG-08,LAU-10}. It represents a cross section
($xy$ plane) of a single crystal with a single slip geometry and
straight parallel edge dislocations along the $z$ axis. The $N$
dislocations glide along directions parallel to their Burgers
vectors $\vec{b}=\pm b \vec{u}_x$. Equal numbers of dislocations
with positive and negative Burgers vectors are assumed, and
dislocation climb is not considered for simplicity. The dislocations
interact through their long-range stress fields,
$\sigma_s(\vec{r}) = Db x(x^2 - y^2)/(x^2+y^2)^2$,
where $D=\mu/2\pi(1-\nu)$, with $\mu$ the shear modulus and $\nu$
the Poisson ratio of the material. The overdamped equations of motion
read $\chi_d^{-1}v_n/b = s_n b [\sum_{m \neq n} s_m\sigma_s(\vec{r}_{nm})
+ \sigma(t) ]$, with $v_n$ the velocity and $s_n$ the sign of the
$n$th dislocation, and $\chi_d$ is the dislocation mobility, implicitly
including effects due to thermal fluctuations. $\sigma(t)$ is the
sinusoidal external stress, $\sigma(t) = \sigma_0 \sin(\omega t)$,
with $\sigma_0$ the amplitude and $\omega$ the angular frequency.
The equations of motion are integrated with an
adaptive step size fifth order Runge-Kutta algorithm, by
measuring lengths in units of $b$, times in units of
$1/(\chi_d Db)$, and stresses in units of $D$, and by imposing
periodic boundary conditions in both $x$ and $y$ directions.
Two dislocations of opposite sign with a mutual distance smaller
than $2b$ are removed from the system (dislocation annihilation).

The simulations are started from a random initial configuration
of $N_0=1600$ dislocations within a square cell of linear size
$L=200b$. The system first relaxes with $\sigma(t) = 0$, to a
metastable dislocation arrangement. After the
annihilations during the relaxation, $N=500-600$ dislocations
remain. Then, the oscillating external stress
is turned on, and the evolution of the system is monitored by
measuring the time dependence of the strain rate,
$d\epsilon (t)/dt \equiv \epsilon_t(t)= b/L^2 \sum_n s_n v_n(t)$.
We simulate the system extensively for a wide range of $\sigma_0$
and $\omega$-values, with several realizations of the random
initial configuration considered in each case.

\begin{figure}[!t]
\begin{center}
\includegraphics[width = 8.0cm,clip]{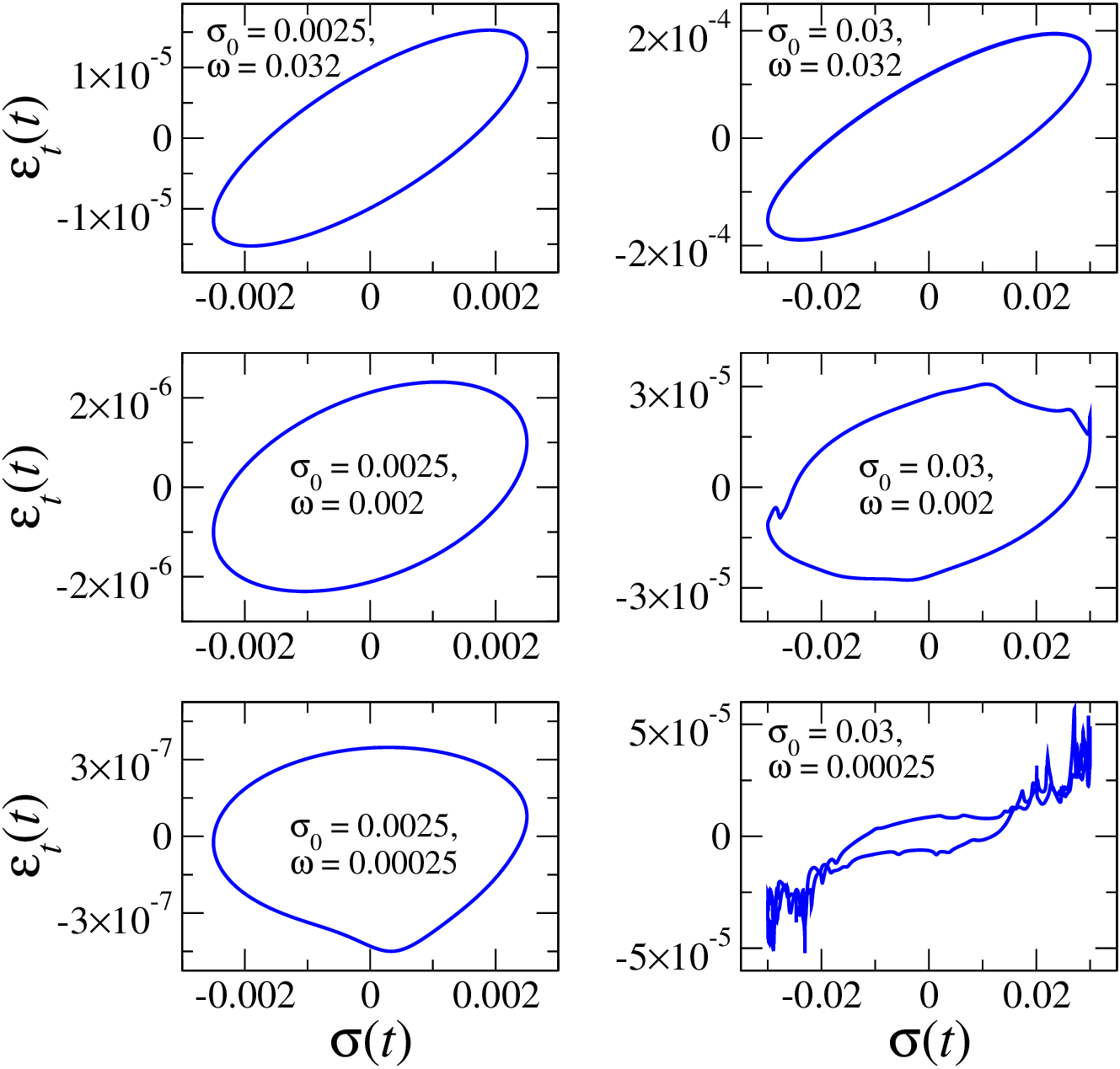}
\end{center}
\caption{(Color online) Examples of typical ``locked-in''
strain rate hysteresis loops reached by the system after a
transient, for low and high stress amplitude (left
and right column, respectively) and for various frequencies.
All loops exhibit a clockwise
rotation direction. For small $\sigma_0$ and/or
large $\omega$ the loops are smooth, but become increasingly
intermittent on increasing $\sigma_0$ and decreasing
$\omega$. Notice that due to the large variation of
$\epsilon_t$-values for the different parameter values
considered, the $\epsilon_t$ axis scales are different in
different sub-figures.}
\label{fig:loops}
\end{figure}

The resulting {\it stress - strain rate} hysteresis loops
exhibit a variety of properties, depending on $\sigma_0$
and $\omega$.
After an initial transient, the system tends to settle into
a ``locked-in'' steady state (usually reached within the 20
cycles we consider) in which the same hysteresis
loop is repeated with a clockwise rotation
direction in the $\sigma$-$\epsilon_t$ plane. Fig.
\ref{fig:loops} shows examples of such locked-in loops for
different $\sigma_0$ and $\omega$. For small $\sigma_0$ and
large $\omega$ (i.e. under conditions where a typical distance
traveled per cycle by a dislocation is small), the loops are
smooth and the strain rate $\epsilon_t(t)$ obeys sinusoidal
dynamics with a well-defined phase difference compared to the
external drive. During the initial transient leading to this
smooth steady state, the system typically exhibits bursty
dislocation rearrangements, but will settle into a smooth
locked-in state after a few cycles. For larger $\sigma_0$
and/or smaller $\omega$, even the steady state cyclic
dislocation dynamics becomes intermittent, characterized by
avalanche-like dislocation rearrangements. Interestingly,
also in this case the system is usually able to find a locked-in
steady state within the 20 cycles we consider, repeating the same
bursty dynamics during each cycle in the steady regime. The
transient time to reach the steady state tends to increase
upon increasing $\sigma_0$ and decreasing $\omega$. For large
$\sigma_0$ and low $\omega$ (bottom right corner of Fig.
\ref{fig:loops}), the loops exhibit curvature consistent with
the idea that the low-frequency limit is described
by $\epsilon_t \sim (\sigma-\sigma_c)^{\beta}$, with
$\beta > 1$ \cite{MIG-02}.

\begin{figure}[!t]
\begin{center}
\includegraphics[width = 7.75cm,clip]{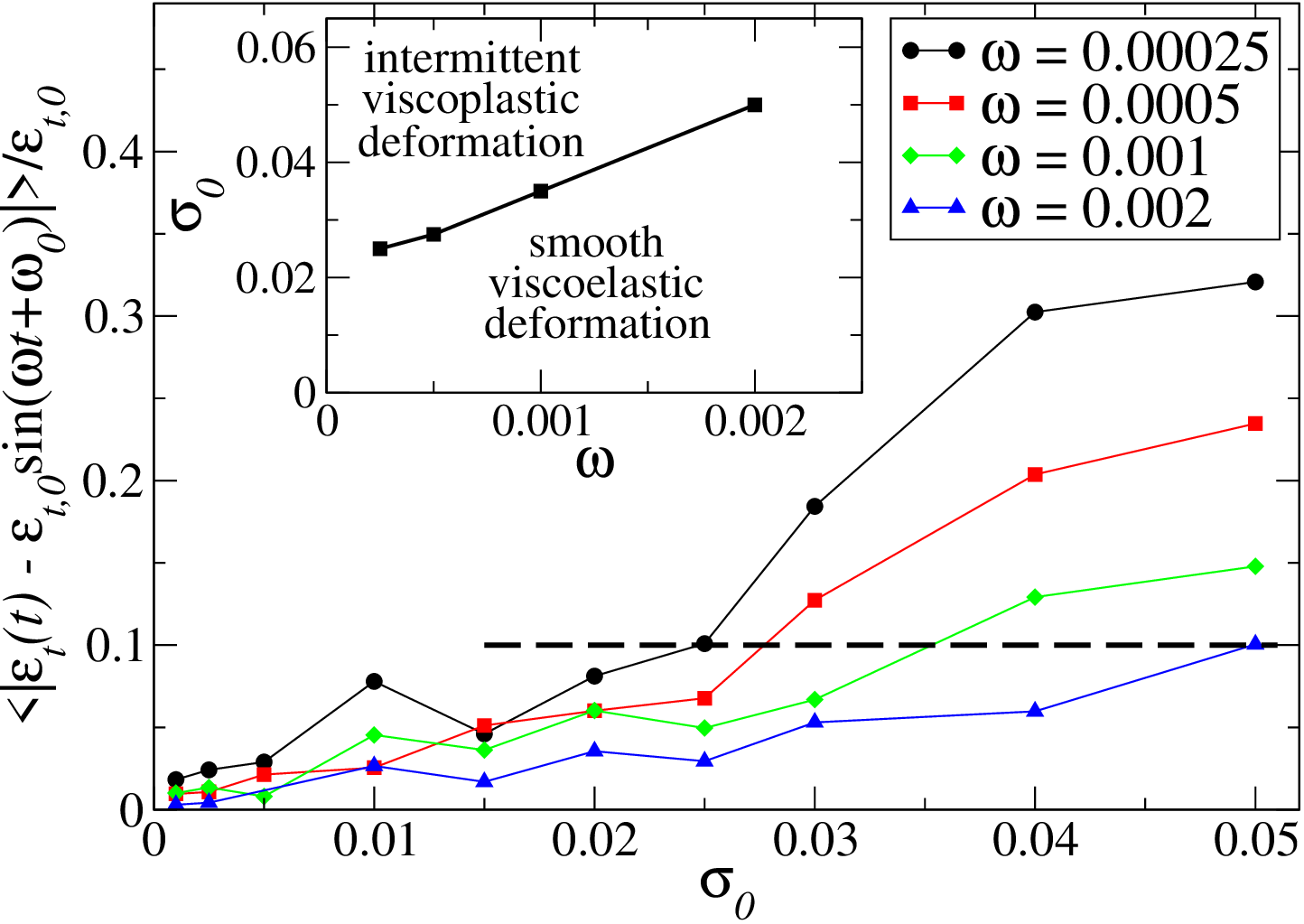}
\end{center}
\caption{(Color online) Normalized absolute deviations
of $\epsilon_t(t)$ from a best-fit sinusoidal function,
$\epsilon_{t,0}\sin (\omega t + \omega_0)$, characterizing the
intermittency of the cyclic dislocation dynamics in
the ``locked-in'' steady state as a function of
$\sigma_0$, for various $\omega$. The inset shows
a phase diagram in the $\omega$-$\sigma_0$ space,
displaying smooth and intermittent phases separated by
a phase boundary obtained by thresholding the data in the
main figure, with the threshold shown as a dashed line.}
\label{fig:intermittency}
\end{figure}

We proceed to characterize the intermittency of the steady
state cyclic dislocation dynamics, by considering the average
normalized absolute deviations of $\epsilon_t(t)$ from a
best-fit sinusoidal function,
\begin{equation}
\Delta \epsilon_t =
\langle |\epsilon_t(t) - \epsilon_{t,0} \sin (\omega t + \omega_0)|\rangle
/ \epsilon_{t,0},
\label{eq:fit}
\end{equation}
where $\epsilon_{t,0}$ and $\omega_0$ are fitting parameters,
and $\langle \dots \rangle$ indicates an average over both
time and different initial configurations. Large values of
$\Delta \epsilon_t$ indicate the presence of non-trivial or
intermittent dynamics. Fig. \ref{fig:intermittency} shows
$\Delta \epsilon_t$ as a function of $\sigma_0$ for various
$\omega$, demonstrating that the intermittency increases with
$\sigma_0$ and decreases with $\omega$. By applying a threshold
value for $\Delta \epsilon_t$, one finds a phase boundary
separating ``phases'' with smooth and intermittent
dynamics in the $\sigma_0 - \omega$ plane (inset of Fig.
\ref{fig:intermittency}). The precise location of
this boundary depends on the threshold value used, but qualitatively
the phase diagram looks the same for a range of threshold values.

\begin{figure}[t!]
\begin{center}
\includegraphics[width = 8.0cm,clip]{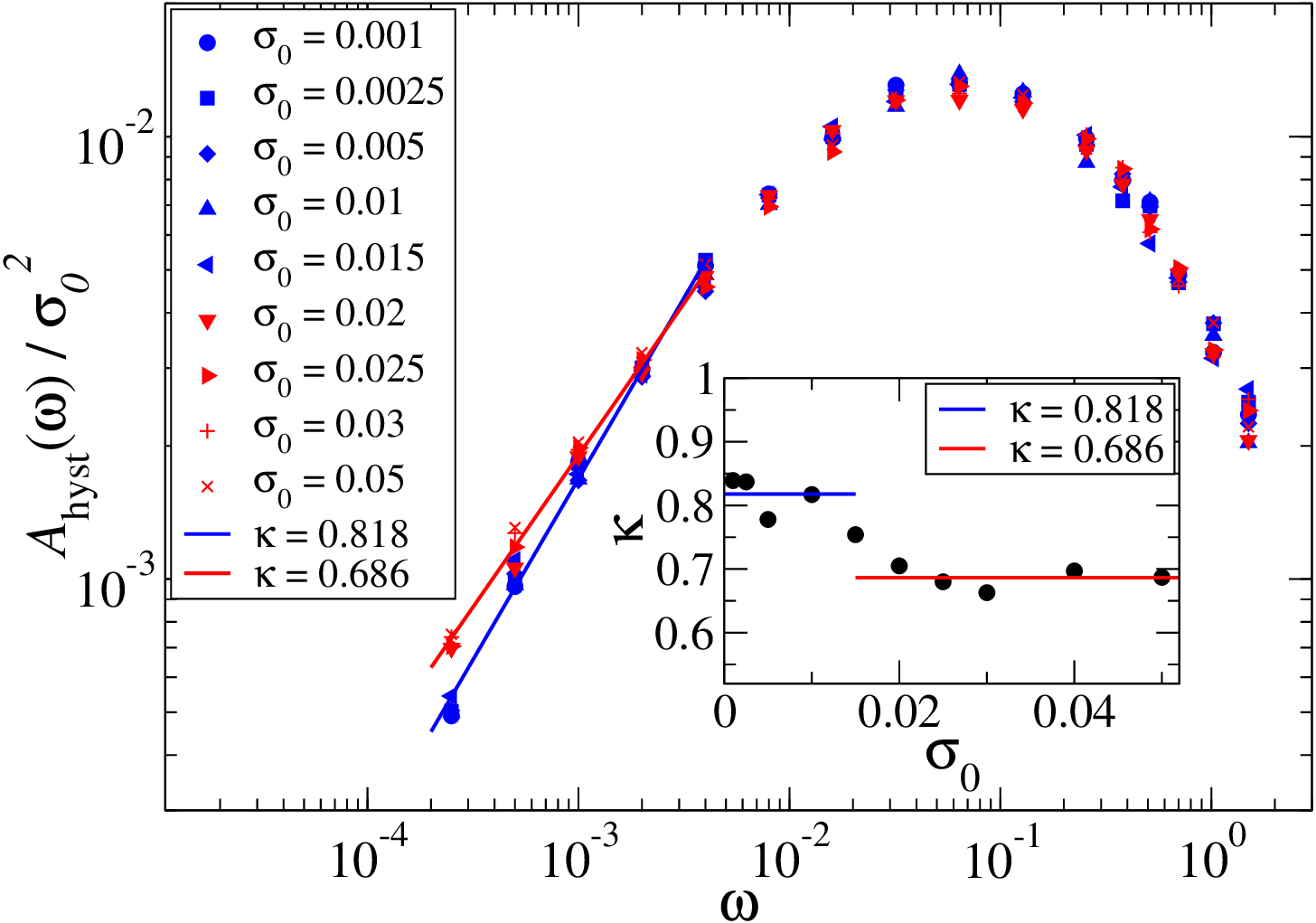}\\
\end{center}
\caption{(Color online) The scaled hysteresis loop areas
$A_{hyst}/\sigma_0^2$ for various $\sigma_0$ as a function of $\omega$,
showing a data collapse with two distinct low-frequency power
laws, $A_{hyst} \sim \omega^{\kappa}$, with $\kappa \approx 0.82$
for $\sigma_0 < 0.015 \approx \sigma_c$ and $\kappa \approx 0.69$
for $\sigma_0 > 0.015$. The inset shows the measured $\kappa$ as
a function of $\sigma_0$.}
\label{fig:areas}
\end{figure}

Our main result concerns the area $A_{hyst}(\sigma_0, \omega)
= \oint \epsilon_t d\sigma$ of the steady state stress - strain
rate hysteresis loops as a function of $\sigma_0$ and $\omega$.
These are summarized in  Fig. \ref{fig:areas}.
$A_{hyst}(\sigma_0, \omega)$
exhibits a maximum at a characteristic frequency
$\omega^* \approx 0.06$ independent of $\sigma_0$, corresponding
to the resonance frequency of the effective confining potential
(see the oscillator model below). The $A_{hyst}(\omega)$-data
for various $\sigma_0$ can be collapsed by normalizing with
$\sigma_0^2$, leading to two distinct low frequency power laws
$A_{hyst} \sim \omega^{\kappa}$, with exponents $\kappa \approx
0.82$ and $\kappa \approx 0.69$ for $\sigma_0 < 0.015$ and
$\sigma_0 > 0.015$, respectively. The stress amplitude value
$\sigma_0 = \sigma_c (N) \approx 0.015$ separating these two
regimes corresponds roughly to the maximum $\sigma$-value for
which the power-law Andrade creep is observed in a
constant stress simulation, i.e. $\epsilon_t \sim t^{-\theta}$
with $\theta$ close to $2/3$ \cite{MIG-02,MIG-08,ROS-10,LAU-11,ISP-11}.
For a larger applied stress in the DC-driven case, the system
would reach a (quasi)stationary moving/flowing state
with a non-zero mean strain rate \cite{MIG-02}.
Thus, we argue that the two stress amplitude regimes with the
different $\kappa$-values correspond for the system to in the jammed 
($\sigma_0 < \sigma_c (N)$) and moving ($\sigma_0 > \sigma_c (N)$) 
states for a constant external stress.

\begin{figure}[t!]
\begin{center}
\includegraphics[width = 8.0cm,clip]{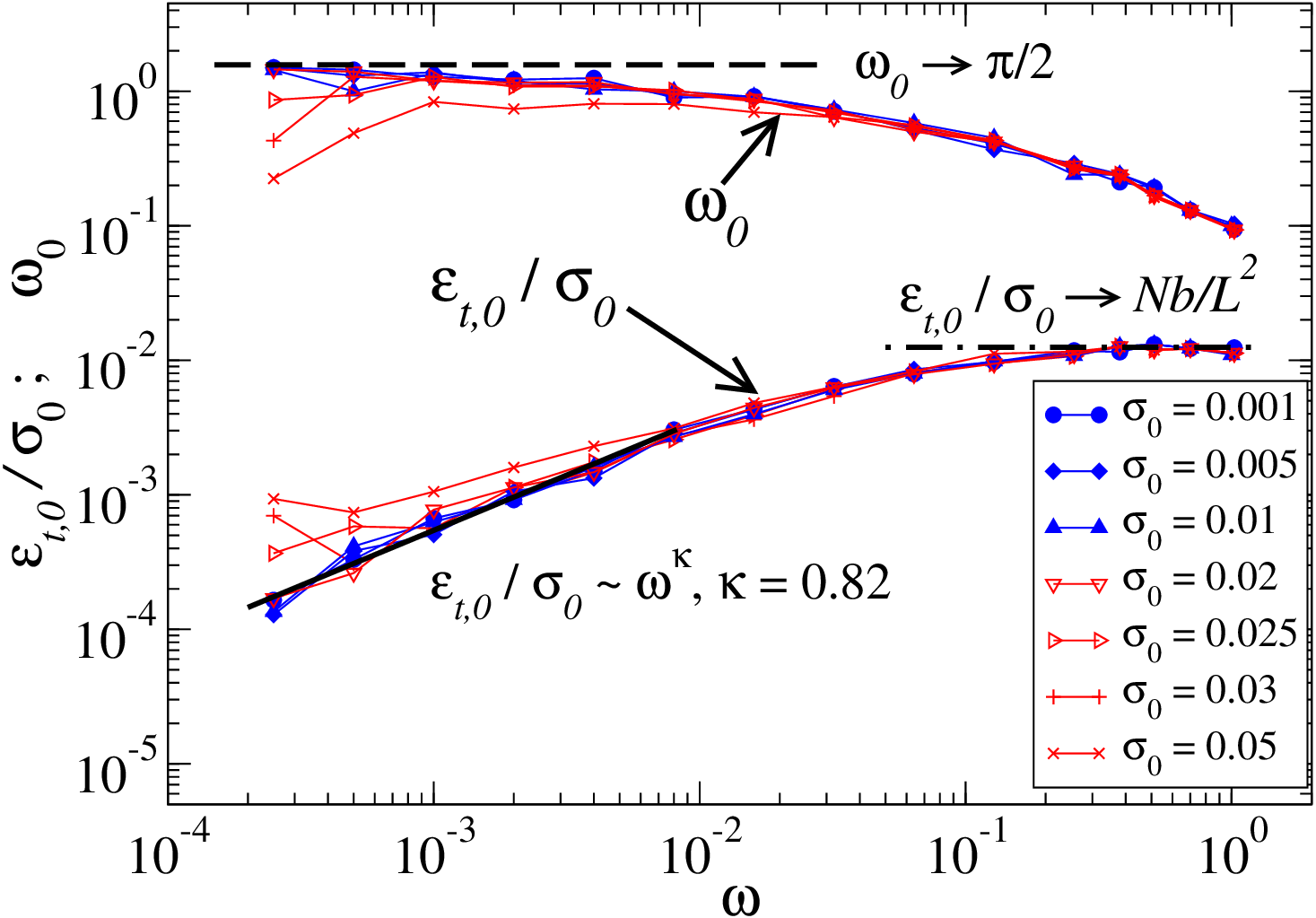}\\
\end{center}
\caption{(Color online) The phase difference $\omega_0$
(upper curves) and the
scaled strain rate amplitude $\epsilon_{t,0}/\sigma_0$
(lower curves) for various $\sigma_0$. For
$\sigma_0 < \sigma_c$, $\omega_0$ approaches $\pi/2$ for
$\omega \rightarrow 0$ (dashed line, a signature of perfectly
elastic dynamics), while for larger $\sigma_0$, $\omega_0$
starts to decrease for small $\omega$, indicating the
presence of plastic rearrangements. $\epsilon_{t,0}/\sigma_0$
obeys a power law $\epsilon_{t,0}/\sigma_0 \sim \omega^{\kappa}$
for $\sigma_0 < \sigma_c$ (solid line, with deviations from the
power law for $\sigma_0 > \sigma_c$), and approaches a value
$\epsilon_{t,0}/\sigma_0 = Nb/L^2 \approx 0.0125$ (dash-dotted
line) for large $\omega$.}
\label{fig:phase}
\end{figure}

We note that the magnitude of
$A_{hyst}$ is related to both the phase difference between $\sigma(t)$ and
$\epsilon_t(t)$, and to the strain rate amplitude, i.e. $\omega_0$
and $\epsilon_{t,0}$ in Eq. (\ref{eq:fit}). Fig. \ref{fig:phase} shows
that for $\sigma_0 < \sigma_c$, $\omega_0$ is independent of $\sigma_0$
and approaches $\pi/2$ for $\omega \rightarrow 0$, and goes to zero for
large $\omega$. Viscoelasticity is typically characterized by the
phase lag $\delta$ between $\sigma(t) = \sigma_0\sin(\omega t + \delta)$
and $\epsilon(t) = \epsilon_0 \sin(\omega t)$, with $\delta = 0$ and $\pi/2$
corresponding to perfectly elastic and viscous dynamics, respectively.
Thus, the relation $\omega_0 = \pi/2 -\delta$ implies that the dynamics
extrapolates between perfect elasticity for $\omega \rightarrow 0$
and perfectly viscous dynamics in the high-frequency limit. For
$\sigma_0 > \sigma_c$, $\omega_0$ starts to
decrease for small $\omega$, indicating the presence of plastic
dislocation rearrangements, also visible in the intermittency of
the dynamics (Fig. \ref{fig:intermittency}). Rescaling the strain
rate amplitude by $\sigma_0$ leads to a data collapse for
$\sigma_0 < \sigma_c$, with $\epsilon_{t,0}/\sigma_0 \sim \omega^{\kappa}$,
$\kappa \approx 0.82$ for small $\omega$, while for large
$\omega$, $\epsilon_{t,0}/\sigma_0 \rightarrow Nb/L^2 \approx 0.0125$,
corresponding to $N \approx 500$ dislocations freely following
$\sigma(t)$ in a system of size $L = 200b$. For $\sigma_0 > \sigma_c$,
there are deviations from the low-frequency power law, again
corresponding to intermittent viscoplastic deformation.

As a naive attempt to understand the observed scaling behavior,
we consider a simple oscillator model for the dislocation
dynamics. In general, dislocations will oscillate back and
forth due to the sinusoidal applied stress. However, dislocation
interactions induce a tendency to form dislocation structures
of varying complexity - {\it dislocation multipoles} - with
each multipole moving together with a strain rate $\epsilon_{t}^{(i)}$
in a way dictated by its net Burgers vector $b^{(i)}$ under
the applied field and interactions with the rest of the system. For
small $\sigma_0$ and large $\omega$ we describe the latter by
a harmonic potential, and write the equation of motion for
the $i$th multipole as
\begin{equation}
\label{eq:ho}
\frac{L^2 \epsilon_{t}^{(i)}(t)}{b_i} \equiv x_{t}^{(i)}(t) =
-K^{(i)} [x^{(i)}(t)-x^{(i)}(0)] + \sigma_0 \sin (\omega t),
\end{equation}
where $K^{(i)}$ is the effective spring constant characterizing
the confining potential of the $i$th multipole (with a center
of mass $x^{(i)}(t)$) due to long-range interactions with the
other multipoles. The asymptotic ($t \rightarrow \infty$) solution
of Eq. (\ref{eq:ho}) is given by
\begin{equation}
\epsilon_{t}^{(i)}(t) =
\frac{b^{(i)} \sigma_0}{L^2}
\frac{[K^{(i)}\omega \cos(\omega t)+\omega^2 \sin(\omega t)]}
{(K^{(i)})^2 + \omega^2}.
\end{equation}
The total strain rate is obtained by summing over the multipoles,
$\epsilon_t = \sum_i \epsilon_{t}^{(i)}$. Disregarding fluctuations by
setting $b^{(i)}=b_{eff}$ and $K^{(i)}=K_{eff}$ for all $i$, one obtains
\begin{equation}
\label{eq:loop}
\epsilon_t(t) =
\frac{b_{eff} N_{mp} \sigma_0}{L^2}\frac{K_{eff}\omega \cos(\omega t)+
\omega^2 \sin(\omega t)}{K_{eff}^2 + \omega^2},
\end{equation}
with $N_{mp}$ the number of dislocation multipoles in the system.
Notice that Eq. (\ref{eq:loop}) corresponds to a clockwise direction
of rotation in the $\sigma - \epsilon_t$ plane, as observed in the
simulations. The area of the hysteresis loop, $A_{hyst}=\oint
\epsilon_t d\sigma $, is given by
\begin{equation}
\label{eq:area}
A_{hyst}(\omega,\sigma_0) =
\frac{b_{eff} N_{mp} \sigma_0^2}{L^2} \frac{\pi K_{eff}\omega}
{K_{eff}^2 + \omega^2}.
\end{equation}
Eq. (\ref{eq:area}) predicts a maximum of $A_{hyst}$ around
$\omega=\omega^*\approx K_{eff}$, and a power law frequency
dependence $A_{hyst} \sim \omega^{-1}$ and $A_{hyst} \sim
\omega^{1}$ for $\omega \gg \omega^*$ and $\omega \ll
\omega^*$, respectively. Fitting Eq. (\ref{eq:area}) to the data
in Fig. \ref{fig:areas} leads to $b_{eff}N_{mp} \approx 400$
independent of $\sigma_0$, suggesting that most dislocations would
move either individually or within wall-like structures. However,
while such a simple model results in the stress amplitude dependence
observed in simulations, i.e. $A_{hyst}(\sigma_0) \sim \sigma_0^2$, it
obviously fails to reproduce correctly the non-trivial
low-frequency $\kappa$-exponents.

Thus it is necessary to go beyond such simplistic descriptions by
considering ideas from critical phenomena, applied to
a yielding transition \cite{MIG-02,LAU-10}. It has been proposed that due to
the long-range dislocation stress fields, this should be described
by the mean-field depinning transition \cite{ZAI-05,DAH-09}. To test
this idea within the present framework, we proceed to contrast our
results with those obtained recently for a mean field elastic
interface subject to AC driving \cite{SCH-11}. There, the exponent
of the low-frequency power-law $A_{hyst}(\omega) \sim
\omega^{\kappa}$ describing the force-velocity (the latter being the
order parameter of the depinning transition) hysteresis loop area
has been shown to exhibit {\it three} regimes as a function of the
applied force amplitude $\sigma_0$, with the exponent $\kappa$
assuming the values $\kappa \approx 0.67$ or 0.75 (for cusped and
smooth disorder, respectively \cite{SCH-11}) for $\sigma_0 \gg
\sigma_c$, $\kappa \approx 0.82$ for $\sigma_0 \approx \sigma_c$ and
$\kappa \approx 1$ for $\sigma_0 \ll \sigma_c$. $\sigma_c$ is the
critical depinning force of the DC-driven system. Thus, in
particular, our numerical results do not agree with the mean field
depinning scaling of the loop area for small force/stress
amplitudes, corresponding to the pinned/jammed phase: The pinned
phase of the mean field interface exhibits trivial dynamics, with
the $\kappa$-exponent coinciding with that of the naive oscillator
model, whereas we find here a $\kappa < 1$ for a wide range of
stress amplitudes with $\sigma_0 < \sigma_c$. An additional
difference is that for dislocations we observe only a single
hysteresis loop \cite{remark}, while for interface depinning models
one typically observes a secondary loop with counterclockwise
rotation direction for $\sigma_0 \gg \sigma_c$ in the region
$\sigma>\sigma_c$ \cite{GLA-03,SCH-11}.

Consequently, our results reveal that various scaling features in
the dynamic hysteresis of crystalline solids relate to the
collective dynamics of dislocations. From the theoretical
point of view the central observation is that unlike the pinned
phase of conventional AC-driven mean field interfaces, the jammed
phase of the dislocation system exhibits critical-like dynamics. In
fact, the $\kappa$-value we find for $\sigma_0 < \sigma_c$ coincides
with the mean field depinning result for $\sigma_0 \approx \sigma_c$
($\kappa \approx 0.82$), suggesting that the system exhibits
criticality in the entire region $0 < \sigma_0 < \sigma_c$. Similar
observations have been made by Isp\'anovity {\it et al.}
\cite{ISP-11}, who found that the dislocation system subject to a
small constant stress exhibits glassy power-law relaxation up to a
time scale limited only by the system size rather than the applied
stress value. We think this is due to the dynamic nature of the
effective disorder - the dislocations are subject and jam due to a
random stress field rather than to quenched disorder absent here,
but fundamental to conventional depinning models. Interesting
extensions of the present study could include considering the effect
of a non-linear mobility law \cite{DISLO,NAD-88,CAI-04} (which
might give rise to a dynamic transition \cite{CHA-99,FUJ-07}), or
cyclic loading of alloys exhibiting the Portevin-Le Chatelier effect
\cite{ANA-07,ZAI-97}.

To conclude, the time-dependent loading of dislocations exhibits
features that result from collective dynamics. This suggests that
such signatures should be seen during the deformation of any
material containing dislocations, and that they should also be
looked-for in the yielding of non-crystalline materials \cite{LEM-09}. 
These findings call for new experimental (for instance in colloidal
crystals \cite{PER-05}) and numerical studies (e.g. molecular
dynamics simulations \cite{MOR-11}) of cyclic dislocation dynamics,
as well as novel theoretical ideas to properly describe dislocation
jamming.

{\bf Acknowledgments}. M.-C. Miguel, M. Zaiser and O. Vartia are
thanked for discussions. This work has been supported by the Academy
of Finland through a Postdoctoral Researcher's Project (LL, project
no. 139132) and via the Centers of Excellence Program (project no. 
251748).


\begin{thebibliography}{10}
\bibitem{remark}
For very large $\sigma_0$ and low $\omega$ we see some double hysteresis,
but in this regime the dislocation displacements are of the order of $L$,
suggesting that periodic boundaries could have an effect.
\end{thebibliography}
\end{document}